\documentclass[aps,floatfix,footinbib,a4paper,prl,superscriptaddress,showpacs,twocolumn]{revtex4}

\usepackage{graphicx}
\usepackage{bm}
\usepackage{amssymb}
\usepackage{color}
\usepackage{amsmath}
\usepackage[normalem]{ulem}

\begin{document}
\title{Noncollinear Andreev reflections in semiconductor nanowires}
\author{B. H. Wu}
\email{bhwu2010@gmail.com}
\affiliation{Department of Applied Physics, Donghua University, 2999 North Renmin Road, Shanghai 201620, China}
\author{W. Yi}
\email{wyiz@ustc.edu.cn}
\affiliation{Key Laboratory of Quantum Information, University of Science and Technology of China,
CAS, Hefei, Anhui, 230026, People's Republic of China}
\affiliation{Synergetic Innovation Center of Quantum Information and Quantum Physics, University of Science and Technology of China, Hefei, Anhui 230026, China}
\author{J. C. Cao}
\affiliation{Key Laboratory of Terahertz Solid-State Technology, Shanghai Institute of Microsystem and Information
Technology, Chinese Academy of Sciences, 865 Changning Road, Shanghai 200050, China}
\author{G.-C. Guo }
\affiliation{Key Laboratory of Quantum Information, University of Science and Technology of China,
CAS, Hefei, Anhui, 230026, People's Republic of China}
\affiliation{Synergetic Innovation Center of Quantum Information and Quantum Physics, University of Science and Technology of China, Hefei, Anhui 230026, China}
\begin{abstract}
We show that noncollinear Andreev reflections can be induced at interfaces of semiconductor nanowires with spin-orbit coupling, Zeeman splitting and proximity-induced superconductivity. In a noncollinear local Andreev reflection, the spin polarizations of the injected and the retro-reflected carriers are typically at an angle which is tunable via system parameters. While in a nonlocal transport, this noncollinearity enables us to identify and block, at different voltage configurations, the noncollinear cross Andreev reflection and the direct charge transfer processes. We demonstrate that the intriguing noncollinearity originates from the spin-dependent coupling between carriers in the lead and the lowest discrete states in the wire, which, for a topological superconducting nanowire, are related to the overlap-induced hybridization of Majorana edge states in a finite system. These interesting phenomena can be observed in semiconductor nanowires of experimentally relevant lengths, and are potentially useful for spintronics.
\end{abstract}
  \maketitle

\emph{Introduction}.--
Andreev reflection (AR) plays a central role in the charge transmission at the interface between a normal conductor (N) and a superconductor (S)~\cite{JETP191228, PRB254515, Tinkham}. An important feature of the process is its spin dependence. For instance, near the interface, an electron with a certain spin injected from N is retro-reflected as a hole of the opposite spin via a conventional local Andreev reflection (LAR). In situations where the spin flip is allowed at the interface, equal-spin AR can occur, where both the electron and the hole have the same spin~\cite{EPL10017012,NP8539}. Recently, it has been shown that equal-spin AR can also be induced by the Majorana bound state (MBS) residing on the edge of a semi-infinite topological superconducting nanowire with spin-orbit coupling (SOC), Zeeman splitting and proximity-induced superconductivity~\cite{PRL112037001,NC53232}. As the equal-spin AR therein originates from the self-Hermitian nature of MBSs, the spin dependence of the AR is bound to be affected in finite-size systems, where overlap coupling between MBSs on different edges invalidates the self-Hermitian property.

\begin{figure}[tbp]
  \centering
  \includegraphics[width=0.6\linewidth]{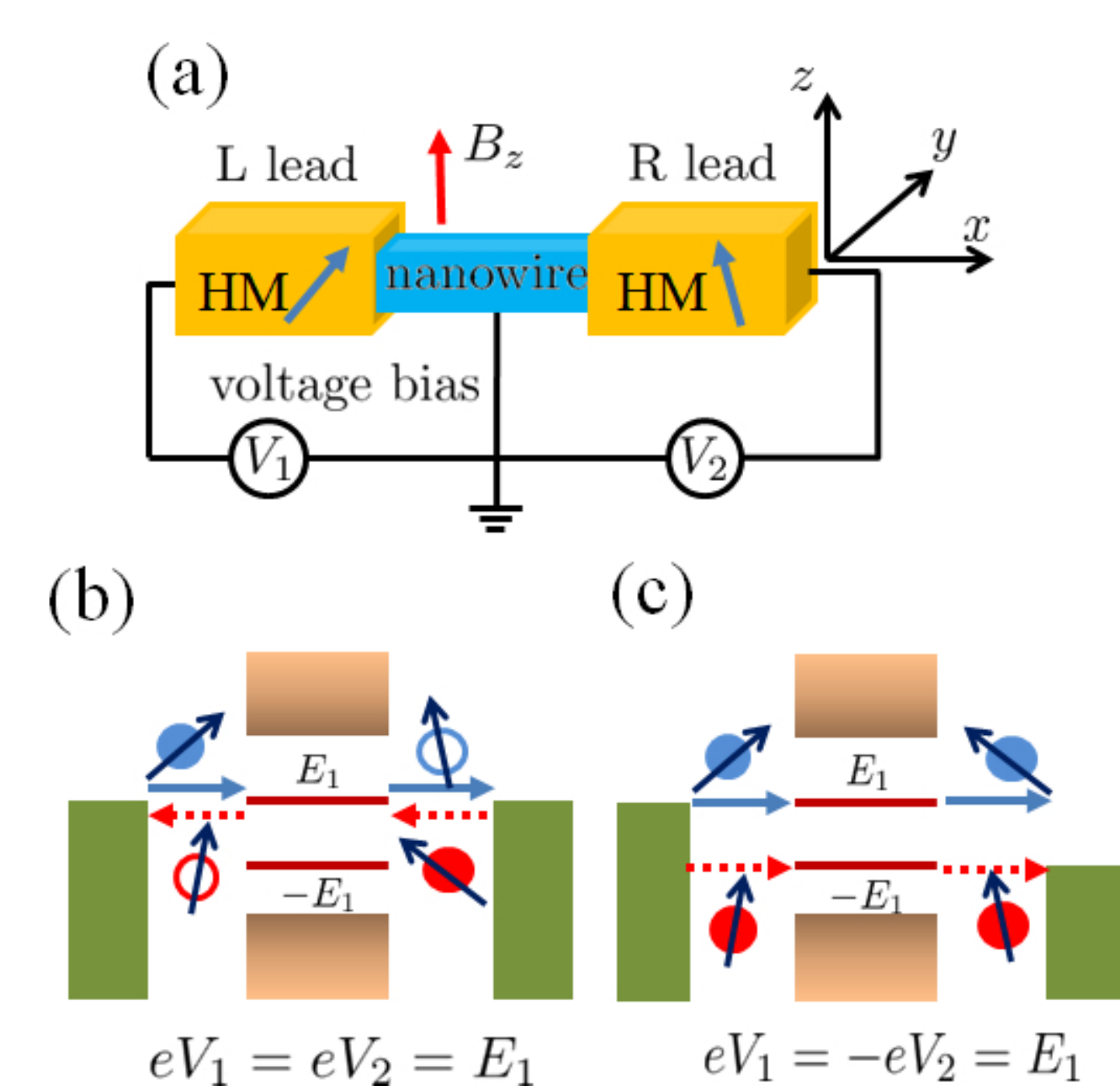}\\
  \caption{(a) A semiconductor nanowire with proximity-induced superconductivity is coupled to two leads. Under different voltage configurations, either a resonant CAR (b) or a direct charge transfer (c) can occur. In either case, two different microscopic transport channels, marked by blue-solid and red-dashed lines respectively, exist, which are characterized by the electrons (filled circle) and holes (empty circle) in the leads with noncollinear spin polarizations (black arrows). The nonlocal transport processes can thus be blocked by adjusting the polarizations of the HM leads, such that transport in both microscopic channels become impossible.}\label{Fig1}
\end{figure}

More generally, for a finite-length nanowire with SOC, Zeeman splitting and proximity-induced superconductivity, pairs of nondegenerate discrete states appear in the energy spectrum~\cite{PRB87024515,PRB87094518}. In a finite-size topological superconducting wire, the lowest discrete states, i.e. the pair of states closest to the Fermi surface, correspond to the in-gap states induced by the hybridization of MBSs on the edges. While in a topologically trivial wire with finite length, the lowest discrete states correspond to the gap-edge states. In this work, we show that carriers involved in a resonant AR with the lowest discrete states can have noncollinear spin orientations, in contrast to the conventional and the equal-spin AR.

In a noncollinear LAR, the spin polarizations of the injected and the retro-reflected carriers are typically at an angle which is tunable via system parameters. Hence, LAR can be suppressed by tuning the alignment of the magnetic moment of the half metal (HM) lead placed at one end of the wire. The noncollinear AR is intimately connected to the local spin orientations of the lowest discrete states at the edge, which are tunable via the Zeeman splitting, the SOC and the wire length. Under appropriate parameters, the conventional or the equal-spin AR can be recovered as limiting cases of the more general noncollinear AR, which is particularly important under realistic experimental conditions of finite-size nanowires.

This noncollinearity is particularly important for nonlocal transport processes where the wire length is typically comparable to the superconducting coherence length~\cite{PRL101120403,PRL103237001,PRB75172503,LS,KZ,PRB88064509,PRL74306,APL76487,PRB6213569,EPJB24287}. In general, when an electron is injected from one N lead to the S, it can either form a Cooper pair in the S and leave an out-going hole in the other N lead; or it can propagate through the S to the other lead. These two different processes respectively correspond to the cross Andreev reflection (CAR)~\cite{PRL101120403,PRL103237001} and the direct charge transfer~\cite{PRB75172503}, which can be differentiated and respectively blocked in a noncollinear nonlocal transport. A paradigm scenario here is by placing two HM leads in contact with the wire (see Fig.~\ref{Fig1}a). Depending on whether the Fermi levels of the two leads are aligned with the same (Fig.~\ref{Fig1}b) or different (Fig.~\ref{Fig1}c) discrete states, either a CAR or a direct charge transfer can take place. Furthermore, by adjusting the polarizations of the HM leads, these noncollinear nonlocal transport processes can be respectively blocked based on the different spin polarizations of carriers in the microscopic transport channels (Fig.~\ref{Fig1}b,c). 
The noncollinearity then provides us with a useful tool to selectively induce fully spin-polarized currents for applications in spintronics.

\emph{Model}.--
We consider a quasi-one-dimensional semiconductor nanowire, which can host MBS for appropriate parameters~\cite{PRL105177002}. As illustrated in Fig.~\ref{Fig1}a, the wire lies in the $x$ direction with an external magnetic field $B_z$ along the $z$ direction. The two ends of the nanowire are tunnel coupled, respectively, to the left (L) and the right (R) leads. The wire is grounded while the two leads are biased by applying voltages $V_1$ and $V_2$, respectively.

In the tight-binding form, the Hamiltonian for the nanowire $H_{\mathrm{wire}}$ reads~\cite{PRB84014503,PRB85085415}
\begin{align}
      H_{\mathrm{wire}}&=\sum_{j=1}^{N-1}\left(-\frac{t_0}{2}c_j^\dag c_{j+1}      -\frac{\alpha_{\rm SO}}{2}ic_j^\dag\sigma_yc_{j+1}+H.c.\right)\nonumber\\+\sum_{j=1}^N&\left[V_zc^\dag_j\sigma_zc_j-(\mu-t_0)c^\dag_jc_j
      +\Delta (c_{j\uparrow}c_{j\downarrow}+H.c.)\right],
\end{align}
where $c_j=(c_{j\uparrow}, c_{j\downarrow})^T$ is the annihilation operator in spinor form for electrons at site $j$ with spin up ($\uparrow$) and down ($\downarrow$) in the $z$ direction, $N$ is the number of the lattice sites, $\sigma_{y,z}$ are the Pauli matrices, $\alpha_{SO}=\alpha_R/a$ is the SOC constant with the Rashba parameter $\alpha_R$ and the lattice spacing $a$, $V_z$ is the effective Zeeman field, $\Delta$ is the proximity-induced superconducting gap, and $\mu$ is the chemical potential. The hopping rate $t_0=\frac{\hbar^2}{m^*a^2}$, with the effective mass of electrons $m^*$. The eigen spectrum of $H_{\mathrm{wire}}$ for a finite-size system consists of pairs of nondegenerate discrete states~\cite{SM}. We may then define the local spin orientation of a discrete state: ${\vec s}_{e/h}(x)=\frac{\langle\psi_{e/h}(x)\mid {\vec \sigma} \mid \psi_{e/h}(x) \rangle}{\langle\psi_{e/h}(x)\mid \psi_{e/h}(x)\rangle}$, which is useful for understanding the noncollinear AR. Here, $\vec \sigma$ is a vector of Pauli matrices and $\psi_{e/h}(x)$ is a spinor in the Nambu space for the electron (hole) component of the discrete state wave function.

The leads are described by the mean-field Stoner model~\cite{PRB65193405}: $H_{\mathrm{lead}}=\sum_{\alpha k s}\epsilon_{\alpha k s}a^\dag_{\alpha k s} a_{\alpha k s}$, where $a_{\alpha k s}$ is the annihilation operator with quantum number $k$, spin index $s$ and energy $\epsilon_{\alpha k s}$ in the $\alpha=$L,R lead. Here, $s=+(-)$ denotes spins parallel (anti-parallel) to the magnetic moment $\bm{n}_\alpha=(\sin\theta_\alpha\cos\varphi_\alpha, \sin\theta_\alpha\sin\varphi_\alpha,\cos\theta_\alpha)$, where $\theta_\alpha$ and $\varphi_\alpha$ are the azimuthal angles in the $\alpha$ lead.

The coupling between the leads and the nanowire is assumed to be spin-conserving: $H_T=\sum_{k}(t_Lc_1^\dag u_La_{L k}+h.c.) +\sum_{k}(t_Rc_N^\dag u_Ra_{R k}+h.c.)$,
where $t_\alpha$ is the hopping between the $\alpha$ lead and the wire,  the electron annihilation operators in spinor form $a_{\alpha k}=(a_{\alpha k +}, a_{\alpha k -})^T$. The unitary matrix $u_\alpha$~\cite{JPCM22035301} accounts for the misalignment of the magnetic moment in the $\alpha$ lead with the $z$ direction
    \begin{eqnarray}
      u_\alpha = \left(
                         \begin{array}{cc}
                           \cos\theta_\alpha/2 & e^{-i\varphi_\alpha}\sin\theta_\alpha/2 \\
                           e^{i\varphi_\alpha}\sin\theta_\alpha/2 & -\cos\theta_\alpha/2 \\
                         \end{array}
      \right).
    \end{eqnarray}
The coupling strength between electrons and the $\alpha$ lead is given by $\Gamma^s_\alpha=2\pi |t_\alpha|^2\rho^s_\alpha$ ($s=\pm$), where $\rho^s_\alpha$ is the density of states in the corresponding lead. In the wide-band limit, $\Gamma^s_\alpha$ becomes energy independent. Under the Stoner model, the spin asymmetry in the leads is characterized by the density of states for the majority ($s=+$) and minority ($s=-$) spins. In particular, for a normal lead, we have $\Gamma^+_\alpha=\Gamma^-_\alpha$; while for fully spin-polarized ferromagnetic or HM leads, $\Gamma^-_\alpha=0$.

We are interested in the transport properties of the system. The current operator for the L lead is related to the evolution of the charge carrier number $N_L=\sum_{ks}a^\dag_{Lks}a_{Lks}$ as $\hat I=-e\dot N_L$~\cite{HJ}. Much information can be obtained from the current fluctuation or noise due to the discrete nature of the charge transport~\cite{PR3361}. Here, we focus on the autocorrelation of the current from the L lead. The zero-frequency noise spectral density $S=\hbar\int dt' \langle \delta I(t')\delta I(0)+\delta I(0)\delta I(t')\rangle$, where $\delta I(t')=\hat I(t)- I$ with $I=\langle \hat I\rangle$ the direct current from the L lead. Both $I$ and $S$ can be evaluated by the standard Keldysh Green's function method~\cite{SM}.

For numerical simulations, we consider a heterostructure in which InSb nanowire is in contact with NbTiN~\cite{Science3361003} or Ni~\cite{NL126414}, with the typical parameters: $m^*=0.015 m_e$, with $m_e$ the electron mass, $\alpha_R=0.25\ \mathrm{eV}\cdot$\AA\ , $a=1$ nm and $\Delta\sim 140$ $\mu$eV. The chemical potential of the wire and the temperature are assumed  zero. With these, we will show that the noncollinear AR can occur in nanowires ranging from submicron to several microns in length.

\begin{figure}
  \centering
  \includegraphics[width=0.65\linewidth]{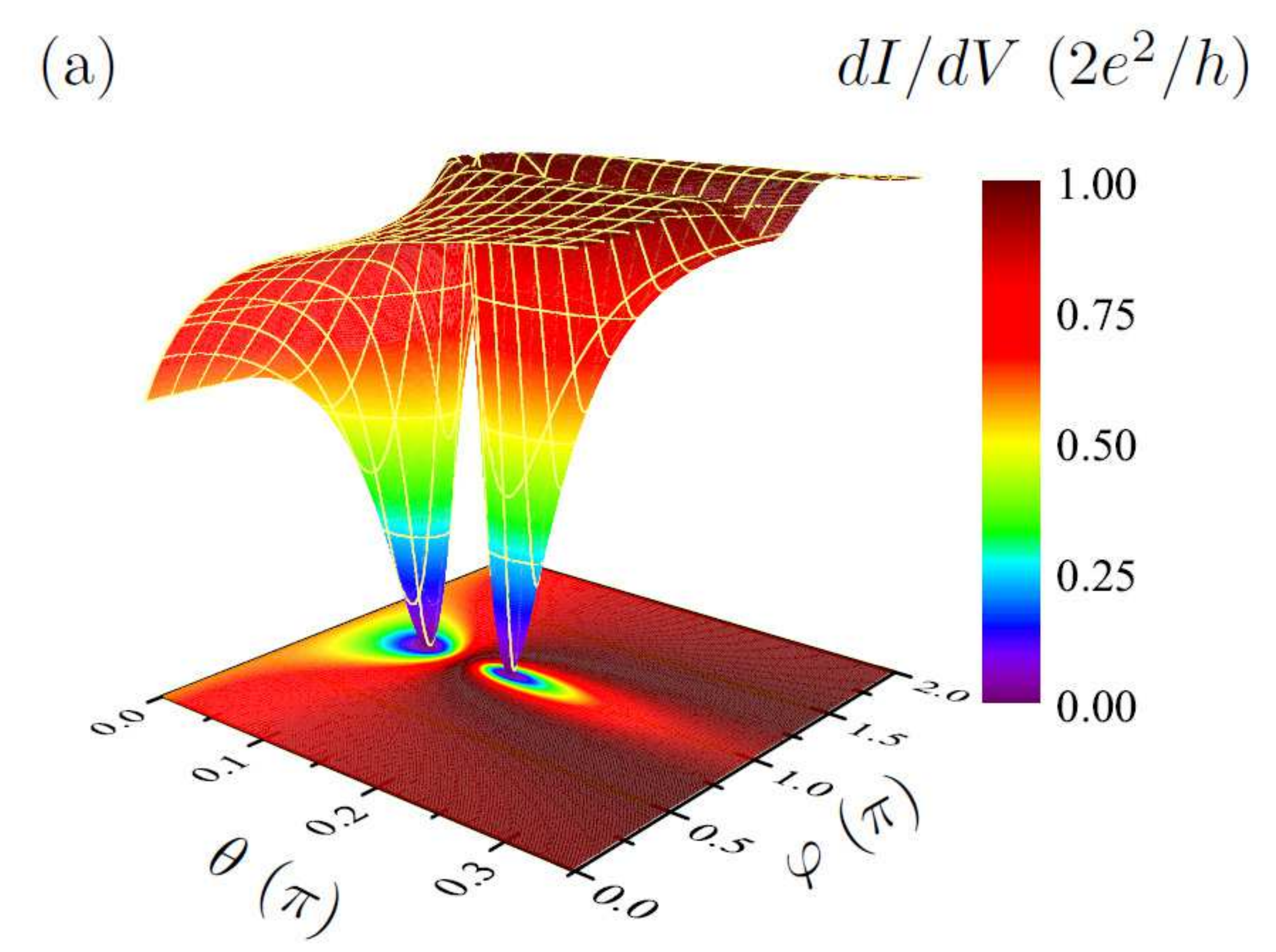}
  \includegraphics[width=0.65\linewidth]{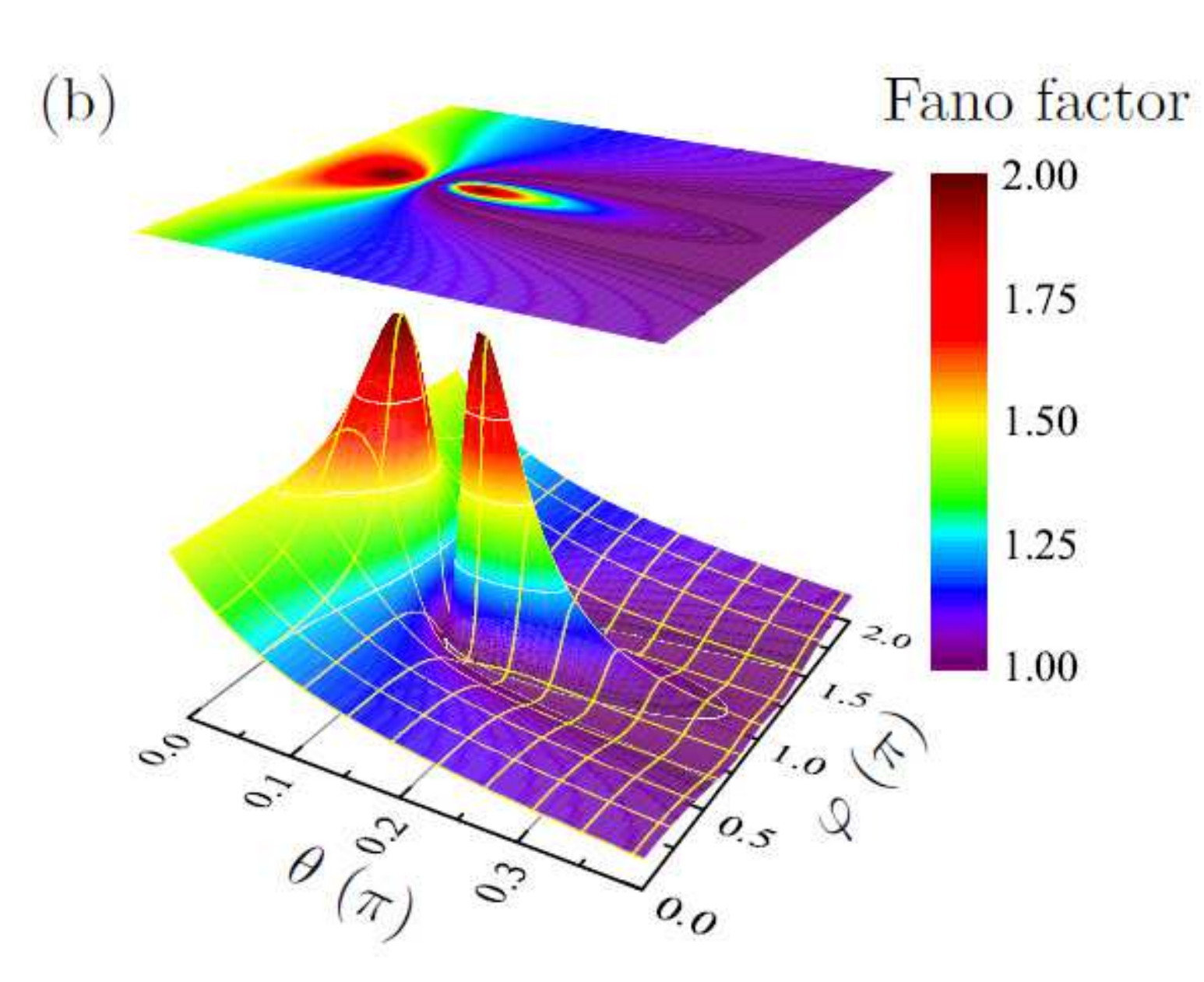}
  \caption{Contour plots of the differential conductance $dI/dV$ (a) and noise Fano factor (b) due to the resonant LAR as functions of azimuthal angles $\theta$ and $\varphi$ of the HM magnetic moment. Parameters: $V_z=1.2\ \Delta$, $\Gamma^+_L=0.4\ t_0$, $L=1.2\ \mu$m.}\label{Fig2}
\end{figure}

\emph{Discrete states and local Andreev reflection}.--
{As the resonant AR is typically connected with discrete states in the wire}, we are interested in the lowest discrete states in a finite system, which, under the particle-hole symmetry, emerge as a pair with the same energy spacing $E_1$ to the Fermi energy. One can show that $E_1$ has nonmonotonic evolution as a function of $V_z$~\cite{PRB87024515,PRB87094518,SM}. These discrete states can be probed by the transport measurement. With an N lead coupled to the end of the wire, the dominant transport process is the LAR.  When the Fermi level of the lead is aligned with $E_1$, conductance peaks of resonant LAR can be identified, where the peak value approaches $2e^2/h$.

The spin dependence of LAR can be identified by replacing the N lead with an HM lead while the bias voltage $V$ satisfies the condition for a resonant AR. We plot the differential conductance $dI/dV$ in Fig.~\ref{Fig2}a as a function of the azimuthal angles $\theta$ and $\varphi$ of the HM magnetic moment. In most situations, $dI/dV$ approaches $2e^2/h$ due to resonant LAR. However, two separate dips can be identified in Fig.~\ref{Fig2}a, indicating the complete suppression of LAR at two particular spin polarizations of the HM lead, as carriers with the desired spin become unavailable in the lead. Since the angle between these two spin polarizations is different from $\pi$ and $0$, Fig.~\ref{Fig2}a clearly indicates a noncollinear LAR. The noncollinear LAR can also be revealed by the current noise Fano factor $F=S/2eI$~\cite{PR3361,PRL763814,PRL824086}. For the LAR-dominated transport, the Fano factor approaches $2$ at low transmission, indicating an effective transfer of two electrons in the process~\cite{PRL103237001}. In Fig.~\ref{Fig2}b, two such peaks emerge in the contour plot of $F$, whose positions are consistent with those of the dips in Fig. 3a.

\begin{figure}
  \centering
  \includegraphics[width=0.650\linewidth]{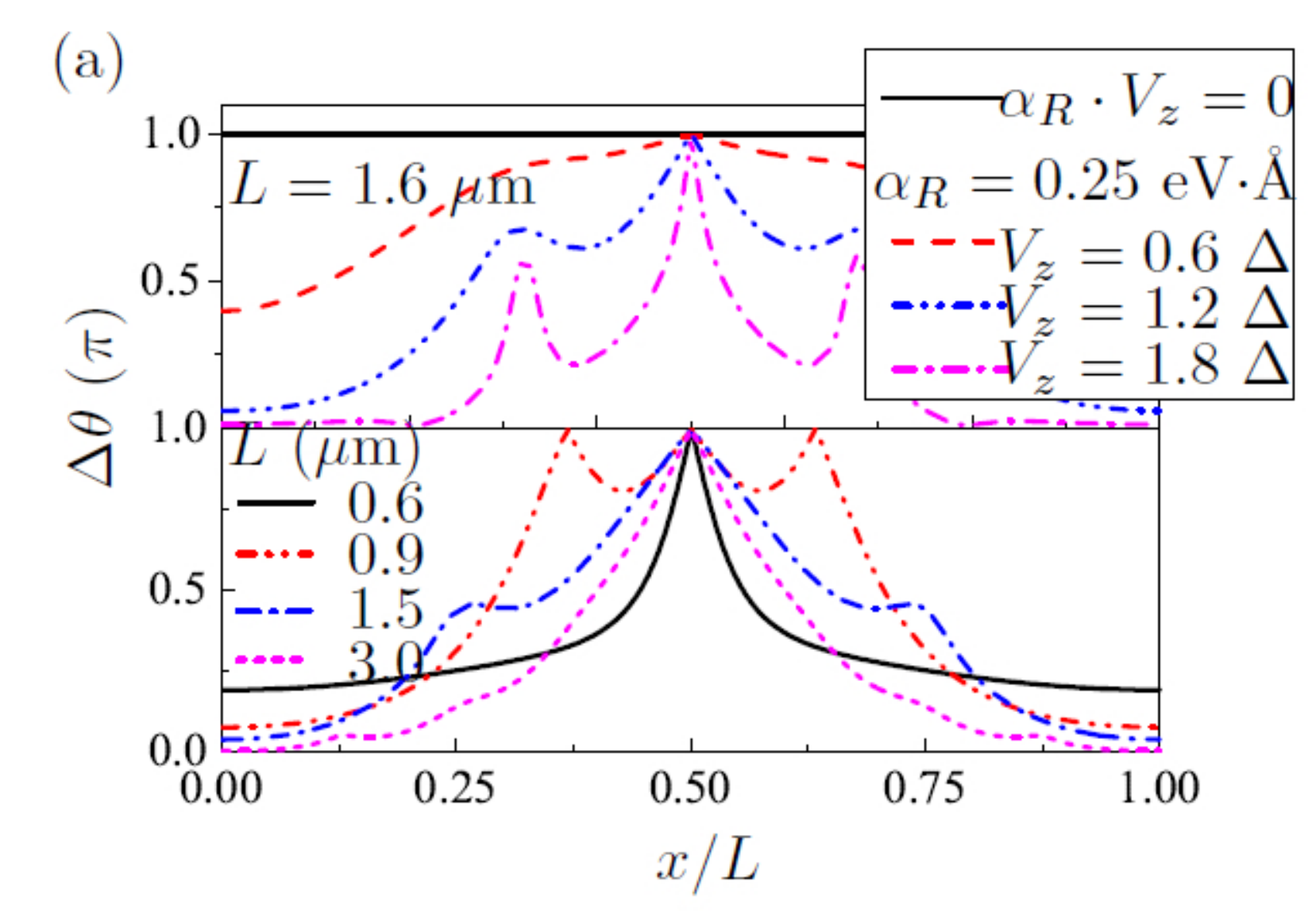}\\
  \includegraphics[width=0.65\linewidth]{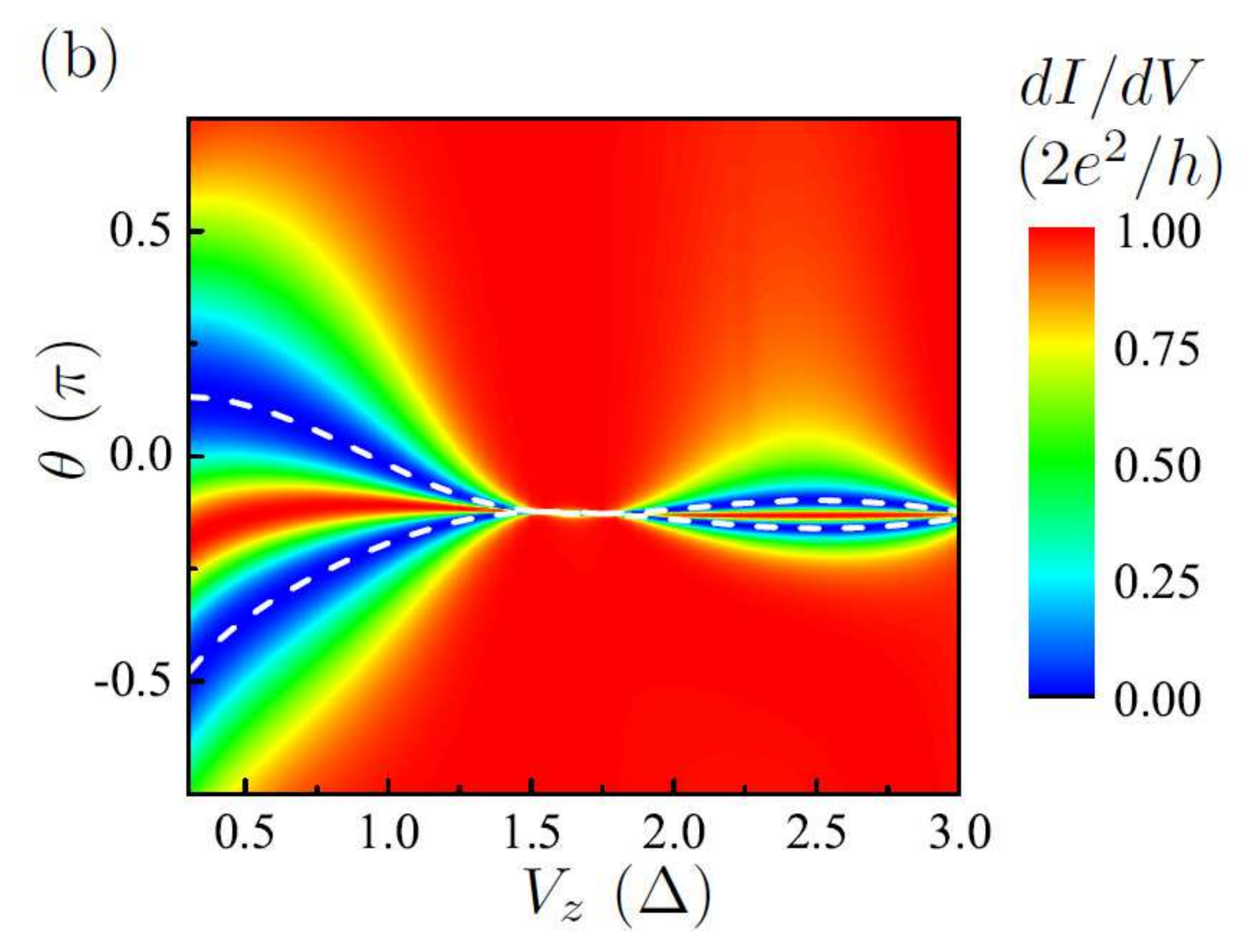}
  \caption{ (a) The spatial dependence of $\Delta\theta$, the angle for the local spin orientations of the lowest discrete state in resonance, for varius $V_z$, $\alpha_R$ and $L$. (b) Contour plot of the differential conductance $dI/dV$ as a function of $V_z$ and the azimuth angle $\theta$ of the magnetic moment in the HM lead. The dashed lines in (b) indicate the directions opposing the local spin orientations of the lowest discrete state on the edge. Other parameters: $\Gamma^+_L=0.4\ t_0$, $V_z=1.2\ \Delta$, $L=1.2\ \mu$m and $\alpha_{R}=0.25\ \mathrm{eV\cdot \AA}$.}\label{Fig3}
\end{figure}

As the coupling between the lead and the wire is spin conserving, it is instructive to explore the local spin orientation of the lowest discrete state in resonance~\cite{PRL096802}. In Fig.~\ref{Fig3}a, we plot the spatial dependence of the angle $\Delta \theta$ between $\vec s_e$ and $\vec s_h$ of the lowest discrete state with energy $E_1$. While the spin orientations of electron and hole components are opposite ($\Delta\theta=\pi$) with vanishing $\alpha_R$ or $V_z$, they become spatially varying and noncollinear under finite SOC parameters, with $\Delta\theta$ at the wire end tunable over a wide range by adjusting $\alpha_R$, $V_z$ or $L$. In particular, in the topological superconducting regime, as $L$ increases, $\Delta\theta$ approaches $0$. This is consistent with the MBS-induced equal-spin AR in Ref.~\cite{PRL112037001}, where the spin orientations of electron and hole components of the MBS are the same due to the self-Hermitian nature of MBS. In a finite topological superconducting wire, the MBSs on different edges overlap to form the lowest discrete states, which acquire a finite $\Delta\theta$. A natural implication here is that the spin-dependence of AR should be intimately connected to the local spin orientation $\Delta\theta$ at the edges.

To see this, we vary the magnetic moment of the HM lead in the $x$-$z$ plane, and adopt the convention: $\theta>0$ for $\varphi=0$; $\theta<0$ for $\varphi=\pi$. Figure~\ref{Fig3}b displays the contour plot of  $dI/dV$ as a function of $\theta$ and $V_z$, for which the voltage $eV=E_1$. The directions opposing that of the local $\vec s_e(0)$ and $\vec s_h(0)$  at the wire end are also displayed with dashed lines. The overlap between the dashed lines and the $dI/dV$ dips clearly indicates that resonant LAR can be suppressed once carriers with spin aligned with either $\vec s_e(0)$ or $\vec s_h(0)$ become unavailable in the lead.

\emph{Crossed Andreev reflection and direct charge transfer}.--
The noncollinearity of the local spin polarizations between the electron and the hole components of the lowest discrete states have significant impacts on the nonlocal transport in nanowires with two leads~\cite{PRL101120403,PRL103237001,PRB88064509, LS, KZ}. Particularly, in a typical noncollinear nonlocal transport, the spin-polarization of the carrier passing through one N-S interface can take an arbitrary angle to that of the carrier at the other interface. Depending on the voltage configuration of the leads, nonlocal transport processes, such as the CAR and the direct charge transfer, can occur, which can be blocked based on their noncollinearity.

We consider the typical case where two HM leads are symmetrically attached to the nanowire. With the wire ends fixed at $x=0$ and $x=L$, the Hamiltonian of the wire observes the symmetry: $H(x,\alpha)=H(L-x, -\alpha)$. For the discrete state with energy $E_1$, the local spin orientations satisfy: $s_{ e/h}^{x,y}(x)=-s_{e/h}^{x,y}(L-x)$ and $s^z_{ e/h}(x)=s^z_{ e/h}(L-x)$.

\begin{figure}
 \includegraphics[width=0.75\linewidth]{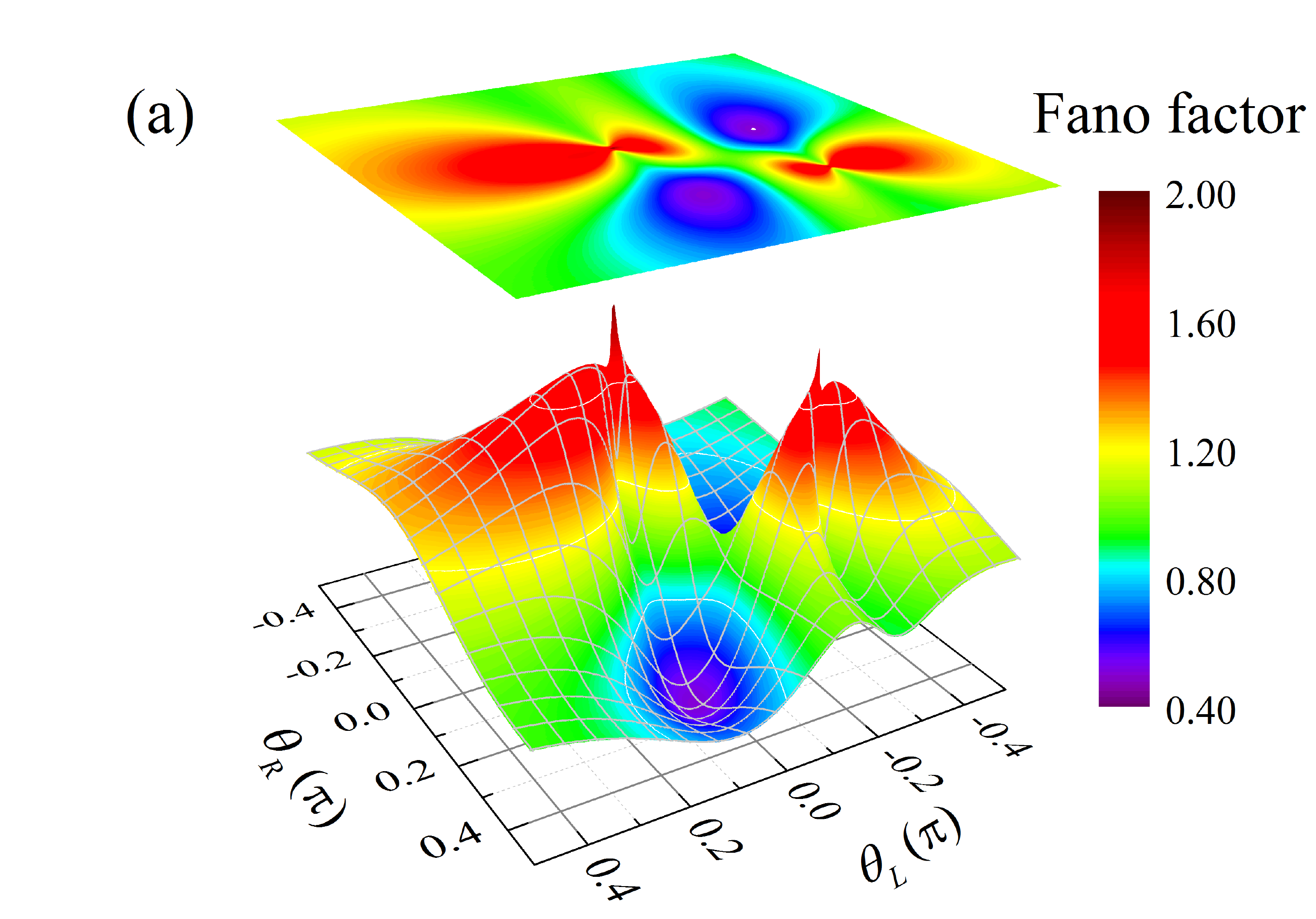}\\
  \includegraphics[width=0.75\linewidth]{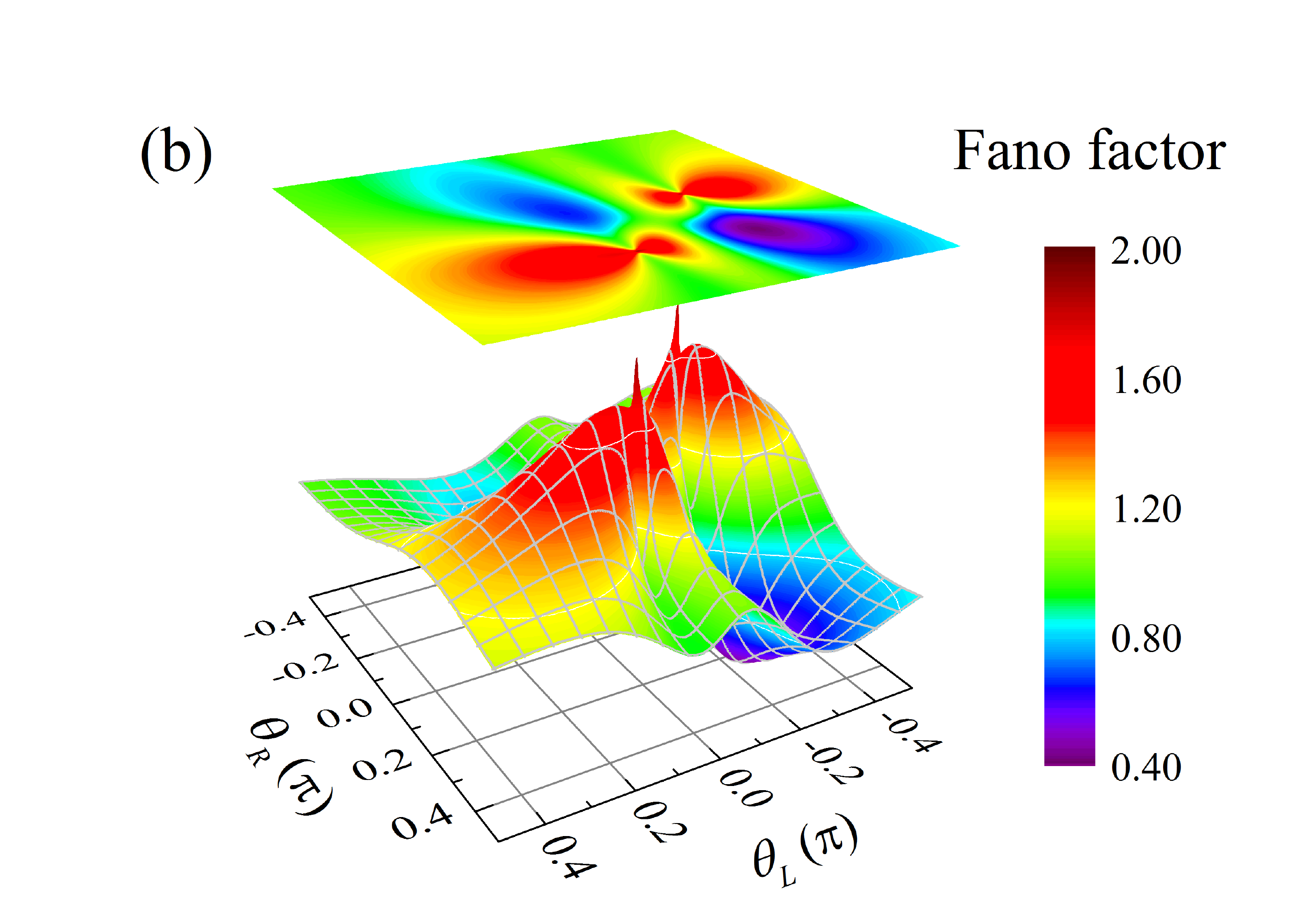}
  \caption{The contour plots of the noise Fano factor $F$ as functions of $\theta_L$ and $\theta_R$ for different bias configurations for the resonant CAR: (a) $eV_1=eV_2=E_1$ and (b) $eV_1=-eV_2=E_1$. Parameters: $V_z=0.8\ \Delta$, $\Gamma^+_\alpha=0.4\ t_0$, $L=1.2\ \mu$m. }\label{Fig4}
\end{figure}

We first consider identical bias voltages $eV_1=eV_2=E_1$, where the Fermi levels of both leads are aligned with the same discrete state (Fig. 1b). The nonlocal transport in this case is a CAR. As demonstrated in Fig.~\ref{Fig4}a, by varying the spin orientations of the L and R leads in the $x$-$z$ plane, two broad humps, each with a sharp feature peaking close to $2$, appear in the noise Fano factor. The peak value $2$ indicates LAR-dominated low transmission, and the suppression of CAR. The locations of both peaks satisfy $\theta_L=-\theta_R$, representing either the simultaneous blocking of carriers with spins aligned in the $\vec{s}_{e}(0)$ direction at the left N-S interface and those with spins in the $\vec{s}_{e}(L)$ direction at the right interface; or the simultaneous blocking of carriers with spin aligned with $\vec{s}_{h}(0)$ at the left and those with $\vec{s}_{h}(L)$ at the right. Here $\vec{s}_{e/h}(x)$ refer to the local spin orientations of the $E_1$ state. The peaks thus indicate the simultaneous blocking of both microscopic transport channels illustrated in Fig. 1b.

Next, we consider the bias voltages $eV_1=-eV_2=E_1$. The nonlocal transport here is the direct charge transfer. The contour plot of the noise Fano factor at the $L$ lead is shown in Fig. \ref{Fig4}b. Again, we find two sharp peaks, with peak values close to $2$. However, as $|\theta_L|\neq |\theta_R|$ at the peaks, they represent either the simultaneous blocking of carriers with spins aligned in the $\vec{s}_{e}(0)$ direction at the left interface and those with $\vec{s}_{h}(L)$ at the right; or the simultaneous blocking of carriers with spins in the $\vec{s}_{h}(0)$ direction at the left and those with $\vec{s}_{e}(L)$ at the right. Again, $\vec{s}_{e/h}(x)$ refer to the local spin orientations of the $E_1$ state, while the particle-hole symmetry dictates that the electron- (hole-) component spin polarizations of the $E_1$ state are identical to the hole- (electron-) component spin polarizations of the $-E_1$ state. The peaks thus indicate the simultaneous blocking of both microscopic transport channels in Fig. 1c.

\emph{Conclusion}.--
We have shown that noncollinear AR can be induced at interfaces of semiconductor nanowires of finite, but experimentally relevant lengths. This feature also provides us with a useful tool to selectively induce fully spin-polarized  current for applications in spintronics. The intriguing effects of noncollinear ARs can be observed in semiconductor nanowires with strong SOC, such as InSb~\cite{Science3361003,NL126414} or InAs~\cite{NP8887,NN979}, which are currently under intensive study for the search of MBS. A test of the noncollinear AR via transport properties is also within the reach of current technology~\cite{PRB85180512}, where supercurrent through HM films such as CrO$_2$ has recently been reported~\cite{Nature439825}.

\emph{Acknowledgement}.--
The authors are grateful for valuable comments from Kam Tuen Law, Xiongjun Liu, Carsten Timm, Philip Brydon, Yuval Ronen. This work is supported by NFRP (2011CB921200, 2011CBA00200), NNSF (60921091), the NSFC (11074266,11105134,11374283) and the Fundamental Research Funds for the Central Universities (WK2470000006, 14D210901).

\newpage

\begin{widetext}

\renewcommand{\theequation}{S\arabic{equation}}
\renewcommand{\thefigure}{S\arabic{figure}}
\appendix
\section{Supplementary material}
In this Supplementary material, we present the formalism for the current and the noise spectral density, as well as the evolution of the energy of the lowest discrete states.

\subsection{Formalism}

 The transport properties such as the differential conductance and current noise can be obtained by the
Green's function method. The current from the L lead is given by
\begin{eqnarray}
      I = \frac{e}{h}\int d\epsilon\mathrm{Re} \mathrm{Tr}\left\{\hat{\bm\sigma}\left[
      {\bm G}^<{\bm\Sigma}^a_{1,1}+{\bm G}^r{\bm \Sigma}_{1,1}^<\right]\right\},
\end{eqnarray}
where $\hat{\bm\sigma}=\mathrm{diag}(1,-1,1,-1)$ in the spin$\otimes$Nambu space accounts for the different charge carried by electrons and holes.
The retarded (lesser) Green's function ${\bm G}^{r/<}$ can be derived from the analytical continuation of the contour-ordered Green's function ${\bm G}(t,t')=-i\langle T\psi(t)\psi^\dag(t')\rangle$, where $\psi_j=(c_{j\uparrow}, c^\dag_{j\downarrow},c_{j\downarrow},c^\dag_{j\uparrow})^T$.
The retarded (advanced) self-energy ${\bm \Sigma}^{r/a}$ with the relation $\bm \Sigma^r=(\bm\Sigma^a)^\dag$
has non-zero elements for lattice sites at the ends of the wire due to the tunnel-coupling. In the wide-band limit, ${\bm \Sigma}^r$ for the outmost sites is given by ${\bm \Sigma}^r_{1,1}=-i{\bm \Gamma}_L/2$ and ${\bm \Sigma}^r_{N,N}=-i{\bm \Gamma}_R/2$, where the matrix $\bm\Gamma_\alpha$ is given in spin$\otimes$Nambu space as
\begin{displaymath}
      \bm{\Gamma}_\alpha=\left(
      \begin{array}{cccc}
        \gamma_{11} & 0 & \gamma_{12} & 0 \\
        0 & \gamma_{22} & 0 & \gamma_{21} \\
        \gamma_{21} & 0 & \gamma_{22} & 0 \\
        0 & \gamma_{12} & 0 & \gamma_{11}
      \end{array}
      \right).
\end{displaymath}
Here, $\gamma_{11}=\Gamma^+_\alpha \cos^2\frac{\theta_\alpha}{2} +\Gamma_\alpha^- \sin^2\frac{\theta_\alpha}{2}$, $\gamma_{22}
    =\Gamma^+_\alpha\sin^2\frac{\theta_\alpha}{2}+\Gamma_\alpha^- \cos^2\frac{\theta_\alpha}{2}$, $\gamma_{12}= (\Gamma^+_\alpha e^{-i\varphi_\alpha}-\Gamma^-_\alpha e^{i\varphi}) \frac{\sin\theta_\alpha}{2}$ and $\gamma_{21}=\gamma_{12}^*$.
The lesser self-energy $\bm{\Sigma}^<$ which characterizes the particle injection from the leads are given by
$\bm{\Sigma}^<=\left[\bm{\Sigma}^a-\bm{\Sigma}^r\right]\bm{F}$, where $\bm{F}_{1,1}=\mathrm{diag}(f(\epsilon-\mu_L),
    f(\epsilon+\mu_L),f(\epsilon-\mu_L),
    f(\epsilon+\mu_L))$ and $\bm{F}_{N,N}= \mathrm{diag}(f(\epsilon-\mu_R),
    f(\epsilon+\mu_R),f(\epsilon-\mu_R),
    f(\epsilon+\mu_R))$ with $f$ the Fermi distribution function.

In this work, we focus on the autocorrelation of the current from the L lead. The zero-frequency noise spectral density is defined as $S=\hbar\int dt' \langle \delta I(t')\delta I(0)+\delta I(0)\delta I(t')\rangle$, where $\delta I(t')=\hat I(t)- I$  is the current fluctuation. $\hat I$ is the current operator. In terms of the Green's functions and self-energies, the noise spectral density can be obtained from
\begin{align}
    S &= \frac{e^2}{h} \int d\epsilon\mathrm{Tr}\left\{\left[\hat{\bm \sigma}{\bm \Sigma}_{1,1}^<\hat {\bm \sigma} \bm G^>+\bm G^< \hat{\bm \sigma}
  \bm \Sigma_{1,1}^>\hat{\bm\sigma}\right]\right.\\
  &\left.-\hat{\bm\sigma}\left[\bm \Sigma_{1,1}\bm G\right]^<\hat{\bm\sigma}\left[\bm \Sigma_{1,1}\bm G\right]^>
  -\left[\bm G\bm \Sigma_{1,1}\right]^<\hat{\bm\sigma}\left[\bm G\bm \Sigma_{1,1}\right]^>\hat{\bm\sigma} \right.\nonumber\\\nonumber
  &\left.+\bm G^>\hat{\bm\sigma}[\bm \Sigma_{1,1}\bm G\bm \Sigma_{1,1}]^<\hat{\bm\sigma}+\hat{\bm\sigma}[\bm \Sigma_{1,1}\bm G\bm \Sigma_{1,1}]^>\hat{\bm\sigma}{\bm G}^<
  \right\},
\end{align}
where the Langreth theorem of analytic continuation such as  $\left[AB\right]^{\lessgtr} = A^rB^\lessgtr
+A^\lessgtr B^a$ and $\left[ABC\right]^\lessgtr=A^rB^rC^\lessgtr+A^rB^\lessgtr C^a +
A^\lessgtr B^aC^a$, have been employed. In the above expressions, the advanced Green's function $\bm G^a=(\bm G^r)^\dag$ and the greater
Green's function $\bm G^>$ can be found from the relation $\bm G^>-\bm G^< = \bm G^r-\bm G^a$.

\subsection{Discrete States}

\begin{figure}
  \centering
    \includegraphics[width=0.5\linewidth]{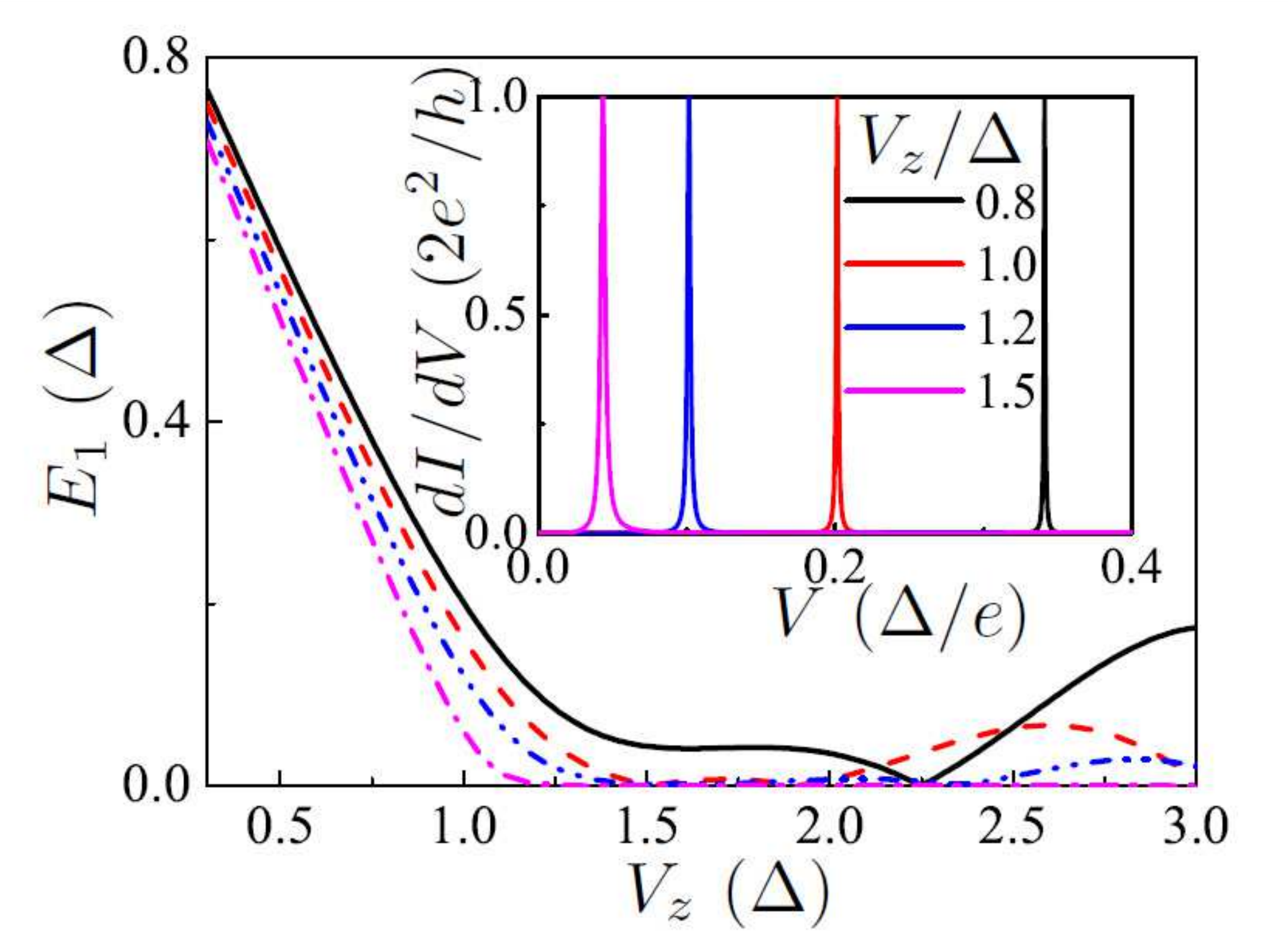}\\
  \caption{The lowest discrete state energy $E_1$
  of the wire as a function of the effective Zeeman field $V_z$ for wire length $L=1.2\ \mu$m (solid line), $L=1.5\ \mu$m (dashed line), $L=2.1\ \mu$m (dash dot dot line) and $L=4.5\ \mu$m (dash dot line).
  The inset: differential conductance $dI/dV$ at various $V_z$. Parameters for the inset: $L=1.2\ \mu$m, $\Gamma_L^{\pm}=0.2 \frac{\hbar^2}{m^*a^2}$ where $a=1$ nm is the lattice space.}\label{Fig2supp}
\end{figure}

The discrete-state energies are obtained by a direct diagonalization of the Hamiltonian (Eq. 1) of the wire in the spin$\otimes$Nambu space.
Due to the particle-hole symmetry, the eigen states of the Hamiltonian appear in pairs that have the same distance to the Fermi surface.
We are interested in the lowest discrete states, i.e., the pair of states that are the closest to the Fermi energy. The nonmonotonic evolution of the discrete-state energy $E_1$ as a function of the Zeeman field $V_z$ for different wire length $L$ is presented in the Fig. \ref{Fig2supp}. In the calculation, we take the following parameters: the pair potential $\Delta \sim 140\ \mu$eV, the spin-orbit coupling $\alpha_R=0.25$ eV$\cdot$\AA\  and the effective electron mass $m^*=0.015\ m_e$, where $m_e$ is the electron mass. The differential conductance is measured when a normal lead is attached to one end of the wire, so that the local Andreev reflection is possible.
The inset shows the differential conductance as a function of the applied voltage $V$ for a wire length of $L=1.2\ \mu$m and with different $V_z$.
The conductance peaks are the clear evidence of the resonant local Andreev reflections due to the discrete states.

\end{widetext}
\end{document}